\documentclass[preprintnumbers, floatfix, twocolumn,
preprintnumbers, PRD, superscriptaddress,nofootinbib]{revtex4-1}
\usepackage{eurosym}
\usepackage{amsfonts}
\usepackage{amsmath}
\usepackage{amssymb,epsf}
\usepackage{latexsym}
\usepackage{graphicx,epsfig}
\usepackage{amssymb}
\usepackage{subfigure}
\usepackage{epstopdf}
\usepackage[colorlinks=true,citecolor=blue,linkcolor=blue,urlcolor=black]{hyperref}
\usepackage{color}
\usepackage{float}

\graphicspath{{Images/}}

\renewcommand{\&}{\textup{\symbol{`\&}}}
\usepackage{hyperref}
\usepackage{amsfonts}
\usepackage{amsmath}

\usepackage{amssymb}
\usepackage{graphicx}%
\usepackage{bigints}
\usepackage{graphicx}
\usepackage{subfigure}
\usepackage{epsf}
\usepackage{bm}
\usepackage{amsmath}
\usepackage{amsfonts}
\usepackage{amssymb}
\usepackage{graphicx}
\usepackage{tabularx}
\usepackage{color}%
\providecommand{\U}[1]{\protect\rule{.1in}{.1in}}

\newcommand{\be}{\begin{equation}}
\newcommand{\ee}{\end{equation}}

\newcommand{\mincir}{\raise
-3.truept\hbox{\rlap{\hbox{$\sim$}}\raise4.truept\hbox{$<$}\ }}
\newcommand{\magcir}{\raise
-3.truept\hbox{\rlap{\hbox{$\sim$}}\raise4.truept\hbox{$>$}\ }}

\providecommand{\U}[1]{\protect\rule{.1in}{.1in}}

\usepackage{tikz,xcolor,hyperref}

\begin{document}

\title{Modified gravity/entropic gravity correspondence due to graviton mass}

\author{Kimet {\bf J}usufi}
\email{kimet.jusufi@unite.edu.mk}

\affiliation{\footnotesize{Physics Department, University of Tetova,
Ilinden Street nn, 1200, Tetovo, North Macedonia}}

\author{Ahmed~Farag~{\bf A}li}
\email{aali29@essex.edu}

\affiliation{\footnotesize{Essex County College, 303 University Ave, Newark, NJ 07102, United States.}}
\affiliation{\footnotesize{Department of Physics, Faculty of Science, Benha University, Benha, 13518, Egypt.}}

\author{ Abdelrahman Yasser }
\email{ayasser@sci.cu.edu.eg}

\affiliation{\footnotesize{Department of Physics, Faculty of Science, Cairo University, Giza 12613, Egypt}}

\author{Nader Inan}
\email{ninan@ucmerced.edu}

\affiliation{ Clovis Community College, 10309 N. Willow, Fresno, CA 93730 USA}
\affiliation{ University of California, Merced, School of Natural Sciences, P.O. Box 2039,
Merced, CA 95344, USA}
\affiliation{ Department of Physics, California State University Fresno, Fresno, CA 93740-8031, USA}

\author{A.Y.Ellithi}
\email{aliellithi@cu.edu.eg}

\affiliation{\footnotesize{Department of Physics, Faculty of Science, Cairo University, Giza 12613, Egypt}}

\begin{abstract}
\noindent
Some time ago, it has been suggested that gravitons can acquire mass in the process of spontaneous symmetry breaking of diffeomorphisms through the condensation of scalar fields [Chamseddine and Mukhanov, JHEP, 2010]. Taking this possibility into account, in the present paper, first we show how the graviton mass intricately reshapes the gravitational potential akin to a Yukawa-like potential at large distances.  Notably, this long-range force modifies the Newton's law in large distances and might explain the phenomena of dark matter. The most important finding in the present paper is the derivation of a modified Newtons law of gravity by modifying the Verlinde's entropic force relation due to the graviton contribution.  The graviton contribution to the entropy basically measures the correlation of graviton and matter fields which then reproduces the Bekenstein-Hawking entropy at the horizon. This result shows the dual description of gravity: in the language of quantum information and entropy the gravity can be viewed as an entropic force, however in terms of particles and fields, it can be viewed as a longe range force. Further we have recovered the corrected Einstein field equations as well as the $\Lambda$CDM where dark matter emerges as an apparent effect. 
\end{abstract}
\maketitle


\section{Introduction}

General Relativity (GR) is considered the most successful physical theory in describing gravity. Its predictions have been tested and confirmed at several length scales. Its gravitational waves have been confirmed and recently observed at \cite{ligo1}. One of the very early attempts to construct a classical relativistic field theory of gravitation was the first N\"{o}rdstrom scalar theory \cite{Nordstrom:1912}. The theory didn't obey the equivalence principle as it insists that the speed of light should be constant in all frames. Such Lorentz covariant gravity theory would imply a dependence between the acceleration of falling bodies and their energies such that the equivalence principle is violated \cite{Norton:2007}. Inspired by the Mach principle, Einstein developed other objections against how N\"{o}rdstrom theory treated rotational motions inconsistently. Eventually, N\"{o}rdstrom suggested modifying his theory such that the source matter should be inertial, and hence, the equivalence principle becomes satisfied \cite{Nordstrom:1913}. N\"{o}rdstrom's second theory was influenced by Einstein-Grossmann \textit{Entwurf} theory, the very early version of the general theory of relativity. It  described gravity by adopting both the metric tensor field and the Laue scalar 
\footnote{Laue scalar is currently known as the trace of the stress-energy tensor.} as sources \cite{Norton:2007}.
When it came to astronomical observations, the new N\"{o}rdstrom theory passed the test of redshift effects. However, it failed to provide a reasonable explanation of the Mercury precession, just like how the Entwurf theory did \cite{Einstein:1913}. But more importantly, Einstein pointed out that despite N\"{o}rdstrom theory being more \textit{natural} than the Entwurf theory, N\"{o}rdstrom theory does not obey Mach's principle, which defines the quality of inertia as a relational physical property. For N\"{o}rdstrom theory to be more natural, Einstein suggested in a 1913 Vienna talk to amend the Entwurf theory by accepting Lorentz invariance, together with the equivalence principle, and by accepting a \textit{relational} definition of gravitational potentials but not absolute ones as in N\"{o}rdstrom theories \cite{Einstein:1913a}. Einstein's strategy relied on adapting a transformation between coordinates using the Jacobian, a method that is now known in the subject of manifold orientation and weight of tensor densities or capacities \cite{Einstein:1913b}. 
General relativity has been incredibly successful in describing gravity on most scales, but there are certain observations, particularly on the scale of galaxies and the universe as a whole, where it seems to fall short. For example, the observed rotation curves of galaxies do not match the predictions of general relativity based on the observed mass distribution, suggesting the existence of unseen "dark matter." Modified gravity theories attempt to explain such phenomena without invoking dark matter or by modifying the behavior of gravity itself. These modifications can take various forms, such as adding new fields to the gravitational sector, modifying the equations of motion, or altering the geometric framework of spacetime itself. Modified gravity therefore are alternatives to Einstein's theory that alters the current description of general relativity at large scales, typically on cosmic scales such as galaxies, clusters of galaxies, or the universe as a whole.

The massive gravity approach suggests that gravitons, the particles believed to mediate gravitational interaction, might have mass. This idea is different from Einstein's theory, where gravitons are massless and move at the speed of light. The concept originated from a seminal study conducted by Fierz and Pauli in the 1930s \cite{FierzPauli1939}. For a considerable period, it was widely believed that the graviton must be massless, owing to a perceived discontinuity between massless and massive theories \cite{Zakharov}. However, this assumption was challenged by Vainshtein \cite{Vainstein}, who demonstrated that below a certain scale, the massive graviton behaves akin to its massless counterpart, suggesting the possibility of a small, nonzero graviton mass. Another significant complication arose in the framework of massive gravity, as highlighted by Deser and Boulware \cite{BoulwareDeser1972}. Their work revealed that alongside the expected five degrees of freedom associated with the massive graviton, an additional scalar degree of freedom emerges, which fails to decouple. Consequently, the theory of massive gravity becomes afflicted with instabilities. Later, de Rham, Gabadadze, and Tolley worked on a version of the theory without issues called \textquotedblleft ghosts\textquotedblright\  \cite{deRham:2011rn}. Another approach for a massive modified gravity theories such as bigravity theories can be ghost-free theories \cite{Hassan:2011zd}. Massive gravity could explain why the universe is expanding at the rate it's observed to, without needing to introduce other things like dark energy \cite{Hinterbichler2012}.
While these theories are free from instabilities mentioned above, there remains a crucial fine-tuning issue. It seems more intuitive to explore the concept of generating mass for gravitons through a spontaneous symmetry breaking mechanism. Such an idea was used recently by Chamseddine and Mukhanov who proposed a Higgs-like mechanism for gravitons \cite{Chamseddine:2010ub}. The theory has no ghosts and the Fierz-Pauli equation is obtained. In this work,  we aim to study the phenomenological aspects of massive gravity and its implications on the dark sector. For example,  important implications of massive gravitons in cosmology were reported recently \cite{J1,J2,J3,J4,J5,J6} where dark matter naturally emerges as an emergent concept. Here we aim to explicitly recover the Yukawa potential and the Modified Newtonian Dynamics (MOND) behavior in large distances. It's worth noting that a related concept, wherein gravitons acquire mass through a spontaneous symmetry breaking mechanism, was proposed in a recent paper of us \cite{J6}. Specifically, the paper demonstrated that gravitons effectively acquire mass within a dark energy superconducting medium. Consequently, the induced mass of gravitons alters both the gravitational potential and Newton's law of gravity over large distances.

Verlinde's proposal to explain the force of gravity as an entropic force opened a new window on the origin of gravity \cite{Verlinde:2010hp}. Further, Verlinde showed the emergence of an extra ''force'' using the
positive dark energy contributes to a thermal volume law
term in entropy. Essentially, Verlinde considered several assumptions to obtain the appearance of the dark
matter force which basically is obtained as an ”elastic” reaction prompted
by entropy displacement dark matter as an emergent effect \cite{Verlinde:2016toy}. In \cite{J7,J8} it was shown the emergence of $\Lambda$CDM using Verlindes emergent gravity. The dark matter is recovered as an apparent effect. Despite the above success, Verlinde's entropic force relation faced criticism from several authors. For instance, potential inconsistencies in Verlinde’s emergent gravity were highlighted in \cite{Dai:2017qkz,Visser:2011jp}. Specifically in \cite{Dai:2017qkz}, it was argued that since gravitational force is conservative, it possesses a fundamental trait common to conservative forces: their actions are inherently reversible. 
This implies that a system in free fall won't inherently increase its entropy, as this process is reversible. To induce entropy growth and irreversibility, some form of dissipation, such as collisions, becomes necessary. Within the framework of general relativity, this dissipation occurs through the emission of gravitons, effectively increasing entropy. Nevertheless, within general relativity, it remains conceivable to devise a freely falling (collapsing) system that doesn't emit gravitons or any other form of radiation, such as a spherically symmetric case. 

Motivated by this, in the present paper we aim to show how the massive graviton can modify the inconsistency. Eventually, as we shall argue, this leads to a modified law of gravity which explains the dark matter problem. The crucial step toward this goal is to add another term due to the contribution of the graviton the the total entropy. In passing, we acknowledge the seminal work by Jacobson \cite{1}, wherein the Einstein equation is essentially derived from the laws of thermodynamics. Additionally, we draw attention to the significant contributions of Padmanabhan \cite{2,3,33}, who argued for the emergence of cosmic space from holographic equipartition. This line of inquiry posits that cosmic space emerges as cosmic time progresses. By quantifying the disparity between surface and bulk degrees of freedom within a spatial region and equating it with the volume change of space, Padmanabhan derived the Friedmann equations describing the universe's evolution. In \cite{J4} the cold dark matter effect was obtained by considering an interacting between the number of degrees of freedom in the bulk. The concepts of entropic force and Padmanabhan's emergent space have garnered considerable attention recently \cite{4,5,6,7,8,9,10,11,12,13,14,15,16,17}, including several works based on a generalised expression for entropy studied in \cite{Nojiri:2022aof,Nojiri:2022dkr,Nojiri:2022nmu,Odintsov:2023vpj}.  Other ideas that include the Bose-Einstein condensate of gravitons to obtain the dark matter effect see \cite{Cadoni:2017evg,Cadoni:2018dnd}.

This paper is organized as follows. In Section II, we review the Chamseddine-Mukhanov mechanism. In Section III, we study  the emergence of a Yukawa potential in GR. In Sections IV, we study the emergence of  dark matter due to a modified gravity force and the emergence of a MOND-like law within GR due to the graviton mass. In Section V, we modify the Verlindes entropic force relation due to graviton mass and we point out a correspondence between the modified Yukawa gravity and entropic gravity. In Section IV, we obtain the entropic corrections to Einstein equations. Finally, in Section VII, we comment on our results. 

\section{Higgs Mechanism for Gravitons: Review of the Chamseddine-Mukhanov model}
In this section, we will review the Chamseddine and Mukhanov model given in \cite{Chamseddine:2010ub} which is based on the following action \cite{Chamseddine:2010ub,Oda:2010wn}
\begin{eqnarray}\notag
S &=& \frac{1}{16 \pi G} \int d^4 x \sqrt{-g} \Biggl\{ R + \frac{m^2}{4} 
\Biggr[ 3\left(\frac{H^2}{16} - 1 \right)^2 
\\
&-& \tilde H_{AB} \tilde H^{AB} \Biggr] \Biggl\}.
\label{CM model}
\end{eqnarray}
These equations exhibit both diffeomorphism and Lorentz invariance and are formulated within the context of 4-dimensional spacetime. Prior to the occurrence of spontaneous symmetry breaking in diffeomorphisms, the system features a massless graviton and a real scalar field $\phi^A$ comprising four components  which gives the induced internal metric 
$H^{AB}$ given by \cite{Chamseddine:2010ub}
\begin{eqnarray}
H^{AB} = g^{\mu\nu} \nabla_\mu \phi^A \nabla_\nu \phi^B,
\label{H}
\end{eqnarray}
where $A = 0,1,2,3$. One can define $H$ and $\tilde H^{AB}$ which are  the trace and the traceless part
of $H^{AB}$, respectively. In particular, one can now define \cite{Chamseddine:2010ub}
\begin{eqnarray}
H^{AB} = \tilde H^{AB} + \frac{1}{4} \eta^{AB} H,
\label{tilde H}
\end{eqnarray}
in which $A, B, \cdots$ are raised and lowered in terms of
the Minkowski metric, $\eta_{AB} = diag(-1, 1, 1,1)$.
From the action, one can find the equations of motion \cite{Chamseddine:2010ub,Oda:2010wn}
\begin{equation}
 \nabla^\mu \Biggl\{ \Biggr[  \tilde H^{AB} - \frac{3}{8} \eta^{AB}
 \left(\frac{H^2}{16} - 1 \right) H \Biggr]  \nabla_\mu  \phi_B \Biggl\}  = 0,
\end{equation}
along with
\begin{equation}\notag
R_{\mu\nu} - \frac{1}{2} g_{\mu\nu} R 
= \frac{m^2}{8} g_{\mu\nu} \Biggr[ 3 \left(\frac{H^2}{16} - 1 \right)^2-\tilde H_{AB} \tilde H^{AB} \Biggr]
\end{equation}
\vspace{-0.5cm}
\begin{equation}
-\frac{m^2}{4} \Biggr[ \frac{3}{4} \left(\frac{H^2}{16}- 1 \right)
H \nabla_\mu \phi^A \nabla_\nu \phi_A-2 \tilde H^{AB} \nabla_\mu \phi_A \nabla_\nu \phi_B \Biggr].
\label{Eq.1}
\end{equation}
At the classical level, it can be verified that the equations presented above admit a solution for the vacuum state: $\phi^A = x^\mu \delta_\mu^A$ and $\left \langle g_{\mu\nu}\right \rangle = \eta_{\mu\nu}$. However, it's important to note that this vacuum solution is not static, primarily due to the presence of the $\phi^0$ component within the $\phi^A$ field. To investigate the dynamics around this vacuum solution, let's expand the fields as follows: 
\begin{eqnarray}
\phi^A = x^\mu \delta_\mu^A + \varphi^A,\qquad\,g_{\mu\nu} &=& \eta_{\mu\nu} + h_{\mu\nu}
\end{eqnarray}
along with
\begin{eqnarray}
H^{AB} = \eta^{AB} - \bar{h}^{AB} + \cdots
\label{H-fluctuation}
\end{eqnarray}
where we have defined $\bar{h}^{AB} \equiv h^{AB} 
- \partial^A \varphi^B - \partial^B \varphi^A$. We further consider a diffeomorphisms in the infinitesimal form that is given by \cite{Chamseddine:2010ub,Oda:2010wn}
\begin{eqnarray}
\delta \varphi^A &=& \xi^\mu \nabla_\mu \phi^A \approx \xi^A,
\nonumber\\
\delta h_{\mu\nu} &=& \nabla_\mu \xi_\nu +  \nabla_\nu \xi_\mu
\approx \partial_\mu \xi_\nu +  \partial_\nu \xi_\mu.
\label{Diffeo}
\end{eqnarray}
Under the influence of diffeomorphisms, it has been established that $\bar{h}^{AB}$ remains invariant. The linearized equations of motion can be expressed as follows \cite{Chamseddine:2010ub, Oda:2010wn}:
\begin{eqnarray}
&{}& \partial^\nu \bar{h}_{\mu\nu} - \partial_\mu \bar{h} = 0,
\end{eqnarray}
and
\begin{eqnarray}\notag
&{}& \Box \bar{h}_{\mu\nu} + \partial_\mu \partial_\nu \bar{h} 
- \partial_\mu \partial_\rho \bar{h}_\nu^\rho
- \partial_\nu \partial_\rho \bar{h}_\mu^\rho - \eta_{\mu\nu} ( \Box \bar{h} \\
&-& \partial_\rho \partial_\sigma \bar{h}^{\rho\sigma}) = m^2 ( \bar{h}_{\mu\nu}
- \eta_{\mu\nu} \bar{h} ),
\label{Linearized Eq.1}
\end{eqnarray}
Upon taking the trace of the aforementioned equation, we arrive at the result $\bar{h} = 0$, and in addition, we can deduce $\partial^\nu \bar{h}_{\mu\nu} = 0.$ Ultimately, through these steps, we can derive the reduced form of Einstein's equations as
\begin{eqnarray}
( \Box - m^2 ) \bar{h}_{\mu\nu} = 0.
\label{Linearized Eq.4}
\end{eqnarray}
This equation exhibits a striking resemblance to the equations of massive gravity initially proposed by Fierz and Pauli \cite{FierzPauli1939}. However, it is worth noting, as highlighted in \cite{Oda:2010wn}, that the equations presented above are expressed in terms of the gauge-invariant quantity $\bar{h}_{\mu\nu}$, which encompasses the scalar fluctuations $\varphi^A$, and accounts for the presence of four degrees of freedom. By employing the definition for $\bar{h}^{AB}$, one can express it as follows, as detailed in \cite{Oda:2010wn}:
\begin{eqnarray}
&{}& \partial^\nu h_{\mu\nu} - \partial_\mu h - \Box \varphi_\mu
+ \partial_\mu \partial_\nu \varphi^\nu = 0,
\end{eqnarray}
and 
\begin{equation}
\Box h_{\mu\nu} + \partial_\mu \partial_\nu h 
- \partial_\mu \partial_\rho h_\nu^\rho
- \partial_\nu \partial_\rho h_\mu^\rho- \eta_{\mu\nu} ( \Box h 
- \partial_\rho \partial_\sigma h^{\rho\sigma}) 
\nonumber
\end{equation}
\vspace{-0.5 cm}
\begin{equation}
=m^2 [ h_{\mu\nu} - \partial_\mu \varphi_\nu - \partial_\nu \varphi_\mu
- \eta_{\mu\nu} ( h - 2 \partial_\rho \varphi^\rho ) ].
\end{equation}
We can now proceed to constrain the remaining diffeomorphisms by imposing gauge conditions, particularly for only three of them, represented by $\varphi_\mu = \partial_\mu \omega$, as detailed in \cite{Oda:2010wn}. Under these conditions, the final equation simplifies to the following form:
\begin{eqnarray}
\partial^\nu h_{\mu\nu} - \partial_\mu h = 0.
\label{Linearized Eq.1-2-1}
\end{eqnarray}
Combining these equations we get 
\begin{equation}
\Box h_{\mu\nu} - \partial_\mu \partial_\nu h 
= m^2 [ h_{\mu\nu} - 2 \partial_\mu \partial_\nu \omega 
- \eta_{\mu\nu} ( h - 2 \Box \omega ) ].
\label{Linearized Eq.1-2-2}
\end{equation}
By performing a trace operation on this equation, one can derive the relationship $h - 2 \Box \omega = 0$, as described in \cite{Oda:2010wn}. With the assistance of Equation (\ref{Linearized Eq.1-2-2}), it can then be deduced that:
\begin{equation}
(\Box - m^2) h_{\mu\nu} = \partial_\mu \partial_\nu \hat{h},
\label{Linearized Eq.1-2-4}
\end{equation}
with the definition $\hat{h} \equiv h - 2 m^2 \omega$. If we further consider the remaining diffeomorphism, $\xi_\mu = \partial_\mu \chi$, one has the following transformations  \cite{Oda:2010wn}:
\begin{eqnarray}
\delta \omega &=& \chi,
\nonumber\\
\delta h_{\mu\nu} &=& 2 \partial_\mu \partial_\nu \chi,
\nonumber\\
\delta h &=& 2 \Box \chi,
\nonumber\\
\delta \hat{h} &=& 2 ( \Box - m^2 ) \chi.
\label{Diffeo remaining}
\end{eqnarray}
By applying this residual diffeomorphism, we can establish the gauge condition $\hat{h} = 0.$ Utilizing this gauge condition, we can deduce that Equation (\ref{Linearized Eq.1-2-4}) transforms into the massive Klein-Gordon equation governing the behavior of $h_{\mu\nu}$. Notably, in the presence of the remaining diffeomorphism, which satisfies the equation $(\Box - m^2)\chi = 0$, it becomes apparent that $\hat{h}$ remains unchanged, as demonstrated in Equation (\ref{Diffeo remaining}). The presence of this residual gauge symmetry enables us to conveniently choose the gauge condition $h = 0$, given that $\delta h = 2 \Box \chi = 2 m^2 \chi$. Consequently, the equations describing the motion of gravitational fields can be expressed in the following form, as outlined in \cite{Oda:2010wn}:
\begin{eqnarray}
( \Box - m^2 ) h_{\mu\nu} &=& 0,
\nonumber\\
\partial^\nu h_{\mu\nu} &=& 0,
\nonumber\\
h &=& 0.
\label{Eq for gravitons}
\end{eqnarray}

Ultimately, it becomes evident that the last equation precisely corresponds to the massive graviton possessing five degrees of freedom, while avoiding the inclusion of the problematic ghost mode, as originally elucidated in \cite{FierzPauli1939}, for more details about 4 scalars see \cite{88,89,90,91}.

\section{Yukawa potential with a massive graviton}

In this section we aim to explore some of the implications of a theory with a massive scalar gravitational potential. Cosmological observations point
out that Einstein's field equation of GR should include a cosmological constant. Then the field equation can be written as%
\begin{equation}
G_{\mu \nu }+\Lambda g_{\mu \nu }=\kappa T_{\mu \nu }^{\text{matter}}
\end{equation}
where $\kappa \equiv 8\pi G/c^{4}$, and $\Lambda $ is the cosmological constant. Again we use $g_{\mu \nu }=\eta _{\mu \nu }+h_{\mu \nu }$, with $%
\left\vert h_{\mu \nu }\right\vert <<1$ in the weak-field limit. Also, the trace-reversed metric perturbation can be defined as
\begin{equation}
\bar{h}_{\mu \nu }\equiv h_{\mu \nu }-\tfrac{1}{2}\eta _{\mu \nu }h,\qquad 
\text{where}\qquad h=\eta ^{\mu \nu }h_{\mu \nu }
\label{Trace-reversed h_mu,nu}
\end{equation}
Using the harmonic gauge, $\partial ^{\nu }\bar{h}_{\mu \nu }=0$, makes the linearized Einstein tensor become%
\begin{equation}
G_{\mu \nu }=-\tfrac{1}{2}\square \bar{h}_{\mu \nu }
\end{equation}
where $\square =\nabla ^{2}-\tfrac{1}{c^{2}} \partial _{t}^{2}$. Letting $T_{\mu \nu }^{\text{matter}}=0$ outside matter sources makes the linearized Einstein field equation become
\begin{equation}
\square \bar{h}_{\mu \nu }-2\Lambda \bar{h}_{\mu \nu }-2\Lambda \left( 1- \tfrac{1}{2}\bar{h}\right) \eta _{\mu \nu }=0  \label{linearized GR with CC}
\end{equation}
When considering gravitational waves, we would use the transverse-traceless field, $h_{ij}^{\text{TT}}$, which satisfies $\eta^{ij}h_{ij}^{\text{TT}}=0$ and $\partial ^{i}h_{ij}^{\text{TT}}=0$. In that case, the entire term involving $\eta _{\mu \nu }$ in $\left( \ref{linearized GR with CC}\right)$ vanishes because we focus on gravitational waves and this term is a constant times the Minkowski metric and gravitational waves depend on $h_{\mu\nu}$, so this term effectively vanishes from the linearized Einstein equation. We get
\begin{equation}
\square h_{ij}^{\text{TT}}-2\Lambda h_{ij}^{\text{TT}}=0
\label{grav wave with CC}
\end{equation}
This is a Klein-Gordon-like equation that matches (\ref{Linearized Eq.4}). Identifying $2\Lambda =m^{2}c^{2}/\hslash ^{2}$ leads to an effective graviton mass given by\footnote{In this section, we have restored the constants $c$ and $\hslash $.}
\begin{equation}
m_{\text{G}}\equiv \frac{\hslash }{c}\sqrt{2\Lambda }\sim 10^{-68}\text{ kg}  \label{m}
\end{equation}%
Also using $\lambda =\hslash /\left( mc\right) $ leads to a length scale given by
\begin{equation}
\lambda_{\text{G}} \equiv \dfrac{1}{\sqrt{2\Lambda }}\sim 10^{26}\text{ m}
\label{lambda}
\end{equation}%
which is consistent with the size of the observable universe. The phenomenological consequences of these results will be explored in a subsequent paper focused specifically on gravitational waves. However, although the results above have been derived for the transverse-traceless field, it will also be shown below to apply similarly to the  Newtonian scalar potential.\bigskip 

In the Newtonian limit, $h_{0i}=0$ and $h_{ij}=h_{00}\eta _{ij}$. We can
define $h_{00}\equiv -\frac{2}{c^{2}}\Phi $, where $\Phi $ is the gravitational scalar potential.
Then $\bar{h}_{00}=\bar{h}=-\frac{4}{c^{2}}\Phi $. Therefore, evaluating $%
\left( \ref{linearized GR with CC}\right) $ for $\left( \mu ,\nu \right)
=\left( 0,0\right) $ leads to%
\begin{equation}
\square \Phi -3\Lambda \Phi =\frac{c^{2}}{2}\Lambda
\end{equation}%
This result matches $\left(\ref{grav wave with CC}\right)$ except for a factor of $3\Lambda $ (instead
of $2\Lambda $), and a source term on the right side. Therefore, the
scalar potential has a corresponding mass and length scale similar to $\left( %
\ref{m}\right) $ and $\left( \ref{lambda}\right) $, respectively, except the factor of $2$
is replaced by $3$. If we introduce a coordinate
transformation to describe the scalar potential due to dark energy as $\Phi
_{D}\equiv \Phi +\frac{c^{2}}{6}$, then the static limit in spherical
coordinates gives%
\begin{equation}
\nabla ^{2}\Phi _{D}=\left( \frac{1}{r^{2}}\partial _{r}\right) \left(
r^{2}\partial _{r}\Phi _{D}\right) -3\Lambda \Phi _{D}=0
\end{equation}%
The solution is a Yukawa potential given by%
\begin{equation}
\Phi _{D}=\frac{c_{1}}{r}e^{-\sqrt{\frac{3}{2}}\frac{mc}{\hslash }r}
\end{equation}%
where $c_{1}$ is a constant determined by boundary conditions. In this form,
it is evident that the gravitational scalar potential has an associated
mass. The usual boundary conditions involving a spherical distribution of
baryonic mass $M$ leads to $c_{1}=-GM\alpha $, where $\alpha $ is a
phenomenological parameter associated with the MOND\ model as will be shown.
Then the Yukawa potential becomes%
\begin{equation}
\Phi _{D}=-\frac{GM\alpha}{r}e^{-r/\lambda }
\end{equation}%
where $\lambda \equiv \lambda _{\text{G}}\sqrt{2/3}$ is the characteristic
length scale associated with the dark energy scalar potential.

One can see that the Newton's constant in a sense is modified $G \to \alpha G$, where $\alpha $ is a phenomenological parameter. The cosmological constant in the field equations is assumed to be nonzero based on observational evidence, and we argued that it could be related to the graviton mass.  On the other hand, if we consider the contribution of the source of matter for the linearized Einstein field equation but we neglect the graviton mass and we set $\Lambda=0$, in that case, we will get in general a spacetime structure that is described by another metric $\sigma_{\mu \nu}$ which is different from $g_{\mu\nu}$. We can call $g_{\mu \nu}$ the physics metric and $\sigma_{\mu \nu}$ the reference metric, respectively.  Working with the approximation $\sigma_{\mu \nu}=\eta_{\mu \nu}+\gamma_{\mu \nu}$, and performing the same analyses we have 
\begin{equation}
\square \bar{\gamma}_{\mu \nu }=-\frac{16 \pi G}{c^4}T_{\mu \nu }^{\text{matter}}.
\end{equation}
If we further assume that matter fields are localised inside some spatial region $\Sigma$, then the solution for the gravitational potential is well known and it is given by
\begin{equation}
\Phi_N =-\frac{GM}{r}.
\end{equation}
Then by superposition, we can find the total Yukawa-like potential as at some large distance 
\begin{equation}\label{Yukawa}
\Phi =-\frac{GM}{r}\left( 1+\alpha\,  e^{-\frac{r}{\lambda }} \right),
\end{equation}
we can see that the total gravitational constant is modified $G\to G+G\alpha=\left( 1+\alpha \right) G$. This analyses hints on the existence of two gravity modes; massive and massless mode and, in a deeper level, probably there exists a more fundamental theory (yet to be found), along with the possibility of having two modes of gravity waves, namely the massless mode along with the massive gravity mode. This is similar to the bimetric gravity theory as was argued for example in Ref. \cite{Aoki:2016zgp} and also extracting bigravity from string theory see Ref. \cite{Lust:2021jps}. We see that in general we have the potential arising in General Relativity due to the spacetime curvature, however a have a Yukawa term due to the long range correlation effect by the massive graviton encoded in the parameters $\alpha$ and $\lambda$.  For the Newtonian case, we know that outside the point particle with mass $M$ we have $\nabla^2 \Phi_N(r)=0$, while in the case of Yukawa corrected potential we have the effective relation $\nabla^2 \Phi(r)=4 \pi G \rho(r)$, where $\rho(r)=-\alpha M e^{-r/\lambda}/(4 \pi r \lambda^2)$ \footnote{Alternatively, we can see the correspondence between the Yukawa-like potential $\Phi(r)=-c_1 e^{-r/\lambda}/r$ and the Newtonian plus Yukawa-like corrections  $\Phi(\tilde{r})=-c_2/\tilde{r}-c_3 e^{-\tilde{r}/\lambda}/\tilde{r}$ in terms of a coordinate change. In the first case, it solves $\nabla^2\Phi(r)=\Phi(r)/\lambda^2=4\pi G \rho(r)$, where $\rho(r)=-c_1 e^{-r/\lambda}/(4 \pi G r \lambda^2).$ In the second case, it solves  $\tilde{\nabla}^2\Phi(\tilde{r})=4 \pi G \rho(\tilde{r})$,  where $\rho(\tilde{r})=-c_3 e^{-\tilde{r}/\lambda}/(4 \pi G \tilde{r} \lambda^2).$ In both cases, one can set $c_1=c_3=\alpha G M$ and obtain the same result for the density which is linked to the dark matter effect. }. This density of matter can be linked to dark matter as we shall elaborate in the next section. It is only an apparent form of matter due to the long range correlation between massive gravitons and matter. In addditon, there is a great level of similarity in the above analyses and the one in \cite{J6} where it it was used  the idea of dark energy medium as a superconductor that basically induces mass to the graviton. The Yukawa-like potential and the role of the graviton mass are also discussed in \cite{Visser:1997hd}. The Yukawa potential has been also used as a way to constrain a fifth force in the ultralight dark sector with asteroidal data \cite{Tsai:2021irw}.

\section{Dark matter: Recovering MOND in massive gravity}
In the upcoming sections, we will explore the interesting aspects of massive gravity phenomenology, with a specific focus on the problem of the dark sector. To begin, (\ref{Yukawa}) can be rephrased as
\begin{equation}
    \Phi=\Phi_B+\Phi_D=-\frac{G M}{r}\left(1+\alpha\, e^{-\frac{r}{\lambda}}\right).
\end{equation}
where we can identify the contribution of the baryonic matter as
\begin{equation}
    \Phi_B=-\frac{G M}{r}
\end{equation}
and the Yukawa correction as an effect due to dark matter,
\begin{equation}
   \Phi_D=-\frac{G M}{r} \alpha\, e^{-\frac{r}{\lambda}}.
\end{equation}
Such Yukawa-like corrections in the potential have been recently linked to the effects of dark matter \cite{J1,J2}.  In the last equation $M$ is baryonic mass and it suggests that dark matter can be viewed as an extra effect due to the modification of the gravitational potential, namely a long-range force of gravity that modifies Newton's law. In other words, there is no dark matter particle needed in this picture. The previous equations hold for a spherically symmetric case. However, in real situations or in cosmological scales, one must define the density of apparent dark matter as an average quantity. Then we can get the rotating flat curves as shown in \cite{J5}. Using the fact that $|F|=m\, v^2/r$, we can rewrite the circular speed of
an orbiting test object as
\begin{equation}
v^2 = \frac{ G M }{r} \left[1+\alpha\,\left(\frac{r+\lambda}{\lambda}\right)e^{-\frac{r}{\lambda}}\right].
\end{equation}
An interesting result is found for galactic scales, leading to a 
MOND-like relation after we rewrite the last equation as
\begin{equation}
\frac{v^2}{r} = \frac{G M }{r^2}+\sqrt{\left(\frac{GM}{r^2}\right)\left(\frac{G M (r+\lambda)^2\alpha^2}{r^2 \lambda^2}\right)e^{-\frac{2r}{\lambda}}}.
\label{speed}
\end{equation}
We can define 
\begin{eqnarray}
    a_B= \frac{GM}{r^2}
\end{eqnarray}
along with
\begin{eqnarray}
    a_0= \frac{G M \alpha (r+\lambda)^2}{r^2 \lambda^2}e^{-\frac{2r}{\lambda}}.
\end{eqnarray}
For the baryonic mass, in general we need to take $M(r)$. 
which can lead to
\begin{eqnarray}
    a=a_B+\sqrt{\alpha\, a_B a_0}.
\end{eqnarray}
Hence, we effectively obtain the famous MOND-like acceleration proposed in \cite{Milgrom1,Milgrom2,Milgrom3}. If we apply our equation for the outer part of the galaxy, we need to get $a_0 \sim 10^{-10} $ m/s$^2$. This is obtained id $\lambda$ is of kpc order to be consistent with the observations $a_0 \simeq 1.2 \times 10^{-10} $ m/s$^2$. Near the central galaxy mass, the first Newtonian acceleration dominates, hence $a\simeq a_B$, however in large distances, the second term proportional to $a_0$ can dominate over the first term $a_B$. This exhibits characteristics similar to MOND and aligns with the findings in the recent paper \cite{Newton–Einstein} that show a gravitational anomaly detected at weak gravitational accelerations in wide binary stars from the Gaia DR3 database.  In real astrophysical situations, we don't have point-like mass distributions, hence one must consider the general expression for the mass distribution. In that case, the apparent dark matter contribution will be modified due to its dependence on the baryonic matter distribution.

Furthermore, the acceleration due to dark matter can be defined as
\begin{equation}
    a_D=\sqrt{\alpha a_B a_0}.
\end{equation}
One can check that for galactic scales where $r \sim \lambda \sim$ kpc, the quantity $a_0$ is very close to the  value for the acceleration of the universe \cite{J5}
\begin{eqnarray}
    a_0 =\lim _{r \to \lambda} \frac{G M  \alpha (r+\lambda)^2}{r^2 \lambda^2} e^{-1}.
\end{eqnarray}

Let us show how the apparent dark matter (the extra contribution of the force) affects the lensing of light. We can use Eq. (\ref{speed}) to obtain the total force on a particle in the case of a galaxy
\begin{equation}
   F =\frac{G M m }{r^2}+\frac{ G M m \alpha (r+\lambda)}{r^2 \lambda} e^{-\frac{r}{\lambda}}=\frac{G (M+M_D) m }{r^2},
\end{equation}
where $M$ is the baryonic matter mass in the galaxy, where we have defined
\begin{eqnarray}
    M_{D}=\frac{ M \alpha (r+\lambda)}{\lambda} e^{-\frac{r}{\lambda}}.
\end{eqnarray}
In the outer part of the galaxy, the baryonic mass can be approximated as a constant term in the limit $r \to \lambda$. For cosmological scales where $ \lambda$ is of Mpc order we need to replace the mass $M$ with the Komar mass $M$ in which case the dark energy contribution will dominate. That is
\begin{equation}
 M =2
\int_V{dV\left(T_{\mu\nu}-\frac{1}{2}Tg_{\mu\nu}\right)u^{\mu}u^{\nu}}.
\end{equation}
where in general we can assume a several matter fluids with a constant equation of state parameters $\omega_i$ satisfying $\dot{\rho}_i+3H(1+ \omega_i) \rho_i=0.$ By taking $ M\sim 10^{52}$ kg, the quantity $a_0$ is very close to the  value for the acceleration of the universe:
\begin{eqnarray}
    a_0 =\lim _{r \to \lambda} \frac{G M (r+\lambda)^2}{r^2 \lambda^2} e^{-1}\sim \frac{G  M}{c^2} \frac{c^2}{\lambda^2} \sim a_{\Lambda}.
\end{eqnarray}
In cosmological scales, dark energy components dominate the total mass, $M=(\rho_\Lambda+3 p_{\Lambda})V$. With $p_{\Lambda}=\omega \rho_{\Lambda}$ ($\omega=-1$), the resulting outward force explains the universe's accelerated expansion. Notably, for the entire universe, $G M/c^2 \sim \lambda $, where $\lambda$ relates to the cosmological constant $\Lambda$. Thus, we can express this as:
\begin{eqnarray}
    a_0 \sim \frac{c^2}{\lambda} \sim c^2 \sqrt{\Lambda}\sim a_{\Lambda}.
\end{eqnarray}
It is therefore very interesting to see that in our picture, one can obtain $a_0 \sim a_{\Lambda}$, where $a_{\Lambda}$ is the acceleration of the universe. This relation it follows naturally in our paper due to the long range force of the massive gravtion. Observational data (\cite{J4,J5}) reveals a discrepancy in the scale of graviton wavelengths: cosmological scales yield $\lambda^{\rm cosmology} \sim 10^{26} \rm{m}$, while galactic scales suggest $\lambda^{\rm galaxy} \sim 10^{19}\rm{m}$. This inconsistency is reconciled by invoking the uncertainty principle for gravitons, $\Delta p \Delta x \sim \hbar$. On cosmological scales, there's greater uncertainty in position but more precision in momentum due to $\Delta p^{\rm cosmology}=m_g c$. This implies a graviton mass of $m_g=10^{-68}\text{kg}$, resulting in $\lambda^{\rm cosmology}=\frac{\hbar}{m_g c}\sim 10^{26} \rm{m}$. Conversely, on galactic scales, there's less uncertainty in position but more in momentum. It's shown (\cite{J3}) that $|\frac{\Delta \lambda}{\lambda}| \sim |\frac{\Delta m_g}{m_g}|$, resolving the inconsistency between measurements. This highlights fundamental measurement limitations, even in cosmology, explaining the discrepancy between galactic and cosmological analyses. The disparity between scales might stem from screening mechanisms like the chameleon mechanism, where the graviton's mass varies based on the surrounding environment.

\section{Relating modified gravity to entropic gravity}
\subsection{Modified second law of Newton}
In this section, we aim to present an important finding: a correlation between the modified law of gravity, specifically Yukawa gravity, and entropic gravity. The concept of entropic force, as introduced by Verinde in his seminal paper \cite{Verlinde:2010hp}, has paved the way for a new perspective on gravitational interaction, framed within the realm of quantum information and entropy. We examine the scenario of a holographic screen and observing its interaction with a particle of mass $m$ as it approaches from the region where spacetime has already manifested. As the particle draws nearer, it gradually integrates with the microscopic degrees of freedom inherent in the screen. According to this view, which is based on arguments given by Bekenstein, it was hypothesized that the alteration in entropy linked to the information on the boundary is equivalent to \cite{Verlinde:2010hp}
\begin{eqnarray}
    {\Delta S} = 2\pi k_B \qquad\quad{\mbox{when}} \qquad\quad
\Delta x = {\hbar \over mc}.
\end{eqnarray}
Note that the inclusion of the factor $2 \pi$ will soon reveal its significance. To provide a slightly more generalized expression, let's rewrite this formula by considering that the change in entropy near the screen follows a linear relationship with the displacement
\begin{eqnarray}
    \Delta S= 2\pi k_B  {mc\over\hbar} \Delta x.
\end{eqnarray}
Verlinde further argues that entropy is proportional to the mass 
$m$, this is justified from the fact that both entropy and mass are additive, it follows naturally that the change in entropy is proportional to the mass. A particle with mass approaches a part of the holographic screen. The screen bounds the emerged part of space, which contains the particle, and stores data that describe the part of space that has not yet emerged, as well as some part of the emerged space. The entropic force in such a case reads \cite{Verlinde:2010hp}
\begin{eqnarray}
     F \Delta x = T{\Delta S}.
\end{eqnarray}
This relationship has undergone critical reevaluation in numerous studies. For instance, potential inconsistencies in Verlinde's emergent gravity were highlighted in \cite{Dai:2017qkz}. As outlined in \cite{Dai:2017qkz}, since gravitational force is conservative, it possesses a fundamental trait shared by conservative forces: their actions are inherently reversible. This means that a system in free fall will not inherently increase its entropy, as this process is reversible. To induce entropy growth and irreversibility, some form of dissipation, such as collisions, is necessary. In the context of general relativity, this dissipation occurs through the emission of gravitons, which effectively increase entropy. Nevertheless, within general relativity, it remains conceivable to devise a freely falling (collapsing) system that does not emit gravitons or any other form of radiation, such as a spherically symmetric case. Motivated by this rationale, our work aims to integrate the influence of massive gravitons, which would consequently modify the entropy. Thus, we propose the following relation
\begin{eqnarray}
    F \Delta x = T\left({\Delta S}+{\Delta S_{\rm graviton}}\right).
\end{eqnarray}
The correction due to the graviton in general can be nonlinear in $\Delta x$ and it includes a correlation between the screen and the particle. Consider the following relation
\begin{equation}
    \Delta S_{\rm graviton}= 2\pi\alpha k_B \frac{m_g c}{\hbar} \Delta x  f(\Delta x),
\end{equation}
where $f(\Delta x)$ is some function which encodes the nonlinear effect of graviton longe range force on the entropy. In Fig. 1, we have shown the plot of $\Delta S_{\rm graviton} / k_B$. This entropy increases and eventually reaches the maximal value at $\Delta x \sim 10^{26}$ which is the radius of the observable universe. This hints on the idea that the Hawking-Bekenstein entropy may be a result of the entanglement entropy due to the correlation between the gravitons and matter fields. In other word, inside the observable radius one has a non-linear law of entropy and at the universe horizon it reproduces the Hawking-Bekenstein entropy. Our results are in agreement with \cite{Bianchi:2012br} where the Hakwing-Bekenstein entanglement entropy was recovered from graviton-matter fields entanglement. In the next section we shall elaborate more on this interesting link.

For the total change in entropy $\Delta \mathcal{S}=\Delta S+\Delta S_{\rm graviton}$, we thus obtain 
\begin{equation}
     \Delta \mathcal{S}=2 \pi k_B \Delta x \left( \frac{m c}{\hbar} + \alpha \frac{m_g c}{\hbar} f(\Delta x)\right).
\end{equation}
In the present paper we shall consider the following choice
\begin{eqnarray}
   f(\Delta x)= \Big[1+ \frac{m_g c}{\hbar} \Delta x+\dots \Big]e^{- \frac{m_g c}{\hbar}\Delta x}
\end{eqnarray}
In this case we define 
\begin{equation}
     \lambda_g =\frac{\hbar}{m_g c}.
\end{equation}
we can write it as 
\begin{equation}
    \Delta S_{\rm graviton}= 2\pi\alpha k_B \frac{\Delta x}{\lambda_g}  \Big[1+\frac{\Delta x}{\lambda_g}+\dots \Big]e^{-\frac{\Delta x}{\lambda_g}}.
\end{equation}

Basically we have a correlation between the particle with mass $m$ and the screen in distance $\Delta x$ and quantified by $\alpha$ due to the graviton mass $m_g$. Thus, in order to have a non zero force, we need to have a non vanishing temperature. From Newton's law we know that a force leads to a non zero acceleration.  Of course, it is well known that acceleration and temperature are closely related. Namely, as Unruh showed, an observer in an accelerated frame experiences a temperature \cite{Verlinde:2010hp}
\begin{equation}
\label{unruh}
 k_BT= {1\over 2\pi} {\hbar\, a_N \over  c},
\end{equation}
where  $a_N$ denotes the acceleration. Let us take this as the temperature associated with the bits on the screen. Further let us use $\Delta x=r$, then one recovers a modified version of second law of Newton
\begin{equation}
\label{secondlaw}
F=m a_N \left(1+\alpha \frac{r+\lambda_g}{\lambda_g}e^{-\frac{r}{\lambda_g}}\right). 
\end{equation}
This revised version of Newton's second law, as described in \cite{Bagchi:2017jfl}, aligns with the generalized uncertainty principle (GUP). Additionally, the modified Newton's second law presented in Eq. (\ref{secondlaw}) seems to correspond to an exponential version of GUP, previously derived in \cite{Miao:2013wua}, which was also employed to explain the value of the cosmological constant. The GUP potentially represents the quantum spacetime as it was first proposed by Snyder \cite{Snyder:1946qz}. In \cite{SU(3)atoms}, It shows a transformative exploration into the universe’s cooling process, detailing how $SU(3) \times SU(2) \times U(1)$ standard model gauge symmetry evolves and simplifies to SU(3) as temperatures plummet towards near-absolute zero Kelvin. This journey, intimately tied to the Meissner effect, posits SU(3) symmetry as the elemental structure of vacuum energy. This paradigm suggests that reaching absolute zero Kelvin is fundamentally unattainable, aligning with the third law of thermodynamics and establishing a novel linkage between this thermodynamic principle and quark confinement. By meticulously quantifying these "atoms" of vacuum energy across the universe, \cite{SU(3)atoms} reveals an exact match with the theoretical to observed vacuum energy density ratio, offering a solution to the cosmological constant problem and defining a mass gap for SU(3) in terms of vacuum energy. This approach of defining a finite number of SU(3) vacuum atoms makes quantum spacetime \cite{Snyder:1946qz} a necessity.

 \begin{figure*}[ht!]
		\centering
	\includegraphics[scale=0.95]{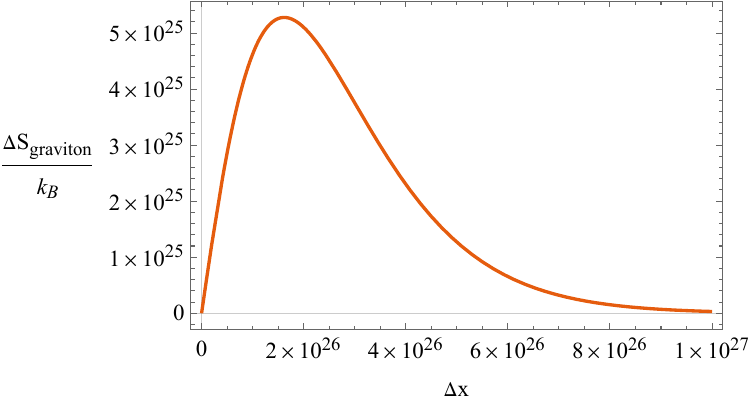}
\caption{Plot of the entropy $\Delta S_{\rm graviton}/k_B$ using $\alpha=0.1 $ and $\lambda_g \sim 10^{26}$ m. The plot shows that this entropy increases and reaches the maximal value at a critical $\Delta x$.  }
	\end{figure*}

In terms of the total acceleration we get 
\begin{equation}
\label{secondlaw}
a=a_N \left(1+\alpha \frac{r+\lambda_g}{\lambda_g}e^{-\frac{r}{\lambda_g}}\right).
\end{equation}
This relation can be written in the MOND-like form as shown in Section  IV. In the limit $\alpha \to 0$, we recover the Verlindes result $F=m a_N$ \cite{Verlinde:2010hp}. Alternatively, one can again consider the Verlindes equation $F \Delta x=T \Delta S$, where in this case $F=m a$, however we need to modify the expression for the temperature 
\begin{equation}
\label{unruh}
 k_BT= {1\over 2\pi} {\hbar\, a \over  c},
\end{equation}
where  $a$ now denotes the modified acceleration. This shows that the acceleration is modified due to the entanglement between the gravitons and matter fields. In both methods, we obtain equivalent results. In the first case we modify the entropy and we get a modified Newton's law, in the second approach we start from a modified Newtons law while the entropy remains unchanged. Previous studies have explored the implications of entropic gravity on photon and graviton masses \cite{MUREIKA_2011}. furthermore, an extension to a 1+1 dimensional framework has been discussed in \cite{Mann_2011}. Notably, these results match the same value derived in these studies.

\subsection{Yukawa modified gravity law}
\textbf{First approach.} Now, our objective shifts towards deriving Newton's law of gravity. Assuming the validity of the holographic principle, the maximum storage capacity, or total number of bits, correlates with the surface area $A$. Indeed, within a theory of emergent space, this area can be defined in a straightforward manner: each fundamental bit occupies one unit cell by definition. Let's designate the number of utilized bits as $N$. It seems reasonable to posit that this quantity will be proportionate to the surface area. Hence, we express this as \cite{Verlinde:2010hp}
\be
\label{bits}
N  
={A c^3\over G\hbar}
\ee 
where we introduced a new constant $G$. Suppose there is a total energy $E$ present in the system.  Let us now just make the simple assumption that the energy is divided evenly over the bits $N$.  The temperature is then determined by the equipartition rule
\be
\label{equipartition}
E = {1\over 2} N k_B T
\ee
as the average energy per bit.
 After this we need only one more equation:
\be
\label{E=Mc^2}
E=M c^2.
\ee
Here, $M$ signifies the mass that would manifest within the region of space enclosed by the screen. Although this mass remains imperceptible directly within the emerged space, its influence is discernible through its energy. Imagine a particle with mass $m$ in proximity to a spherical holographic screen. The energy disperses uniformly across the occupied bits, mirroring the mass $M$ that would materialize within the space encompassed by the screen. Utilizing the equation
$$A=4\pi R^2$$ along with
\be
\label{entropic}
 F \Delta x = \frac{ E G \hbar }{2 \pi k_B R^2 c^3}\left({\Delta S}+{\Delta S_{\rm graviton}}\right).
\ee
and one obtains the Yukawa modified law of gravity
\be
F={GMm\over R^2}\left(1+\alpha \frac{R+\lambda_g}{\lambda_g}e^{-\frac{R}{\lambda_g}}\right).
\ee

The last relation shows that the Yukawa modified law of gravity is a consequence of the modified entropy and the presence of gravitons.  In this way we will provide a duality that in the language of information and entropy gravity can be thought of as an entropic force, however, in terms of particles and fields it is a longe range force. 

\textbf{Second approach.} However let us see another way of obtaining the same result. Namely, if we define the following dark matter mass 
\begin{eqnarray}
    M_{D}:= \alpha M \frac{R+\lambda_g}{\lambda_g}e^{-\frac{R}{\lambda_g}}
\end{eqnarray}
Then the total mass enclosed the surface with radius $R$ will be $M_{\rm total}=M+M_D$. In such a case, we can write the law of gravity as a square law. To see this,  we need to modify the total energy as
\be
\label{E=Mc^2}
E=M_{\rm total} c^2=(M+M_D)c^2.
\ee
while the entropic force relation can be written without the corrections but just as Verlindes suggested
\be
\label{entropic}
 F \Delta x = \frac{ E G \hbar }{2 \pi N k_B R^2 c^3} \Delta S
\ee
yielding again the saw expression for the modified force
\begin{eqnarray}
   F={GMm\over R^2}\left(1+\alpha \frac{R+\lambda_g}{\lambda_g}e^{-\frac{R}{\lambda_g}}\right).
\end{eqnarray}

This shows one can obtain the dark matter effect either by modifying the entropy, or by modifying the total energy. In the limit $\alpha \to 0$, we obtain the Verlindes result for gravity
\be
F={GMm\over R^2}
\ee

In order to explain the emergence of the dark matter effect \cite{Verlinde:2016toy}, Verlinde incorporated additional assumptions and insights from quantum information theory, where positive dark energy contributes to a thermal volume law term in entropy. Essentially, the appearance of the dark matter force stems from an "elastic" reaction prompted by entropy displacement. However, in our current study, we have demonstrated that no additional assumptions are necessary beyond the interaction of gravitons with matter fields. Remarkably, the dark matter force arises organically due to the graviton's mass governed by a Yukawa-like relationship.

\section{Entropy corrections to Einstein equations}
\subsection{Recovering the Hawking-Bekenstein entropy}
Another important point in Verlinde's derivation of Newton's law of gravitation is the entropy-area relationship $S=A/4$ of
black holes in Einstein's gravity, where $A =4\pi R^2$ represents
the area of the horizon. In Verlinde's theory one has a volume law of entanglement entropy due to the dark energy and, at the horizon, this repoduces the Hawking-Bekenstein entropy. In our case however, we consider the graviton contribution which as we saw modifies the law of gravity and here we would like to point further arguments about the possible role of gravitons in explaining the Hawking-Bekenstein entropy. In other words, can one recovers the Hawking-Bekentesin entropy from the entanglement of gravitons with matter field and dark energy. To shed light on this let us assume the following modification \cite{J1,J2,17} 
\begin{equation}
 S=\frac{A}{4}+\mathcal{S}(A). 
 \end{equation}

The standard Bekenstein-Hawking result is reproduced when the second term vanishes. The entropy of the surface changes by one fundamental unit $\triangle S$ fixed by the discrete spectrum of
the area of the surface via the relation 
\begin{equation} 
d S=\frac{\partial S}{\partial A}d A=\left[\frac{1}{4}+\frac{\partial \mathcal{S}}{\partial A}\right]d A. 
 \end{equation}
Here we note that the energy of the surface $\Sigma$ is identified with the
relativistic rest mass $M$ of the source mass, $E=M$. On the surface $\Sigma$, we can relate the area of the surface to the number of bytes according to $ A=QN,$
where $Q$ is a fundamental
constant, and $N$ is the number of bytes. Let us assume that the temperature
on the surface is $T$, by means of the equipartition law
of energy, we get the total energy on the surface via 
\begin{equation}
   E=\frac{1}{2}Nk_B T.
\end{equation}
Further taking $\triangle N=1$,and $\triangle A=Q$, one can get (see, \cite{17})
\begin{equation}\label{F4}
F=-\frac{GMm}{R^2}\left[1+4\,\frac{\partial \mathcal{S}}{\partial A}\right]_{A=4\pi R^2}.
\end{equation}

Considering the Yukawa-type gravitational potential and by using the relation $F=-\nabla \Phi(r)|_{r=R}$, we obtain the modified  a modified Newton's law
\begin{equation}
F=- \frac{ G M m }{R^2} \left[1+\alpha\,\left(\frac{R+\lambda}{\lambda}\right)e^{-\frac{R} {\lambda}}\right].
\end{equation}

Thus, with the  correction
in the entropy expression, we see that \cite{J1}
\begin{equation}
1+\left(\frac{1}{2 \pi R}\right)\frac{d \mathcal{S}}{dR}=\left[1+\alpha\,\left(\frac{R+\lambda}{\lambda}\right)e^{-\frac{R}{\lambda}}\right].
\end{equation}
Solving for entropy  we obtain
\begin{equation}\label{entropy1}
\mathcal{S}=\mathcal{C}-2 \pi  \alpha\left(R^2+3\lambda R+3\lambda^2 \right) e^{-\frac{R}{\lambda}},
\end{equation}
where $\mathcal{C}$ is a constant of integration. This means that the total entropy is given by
\begin{eqnarray}
    S=\frac{A}{4}+\mathcal{C}-2 \pi  \alpha\left(R^2+3\lambda R+3\lambda^2 \right) e^{-\frac{R}{\lambda}},
\end{eqnarray}
In passing we point out that $ \mathcal{C}$ can encode corrections related to the space-time itself as we are going to argue bellow. In general the constant $\mathcal{C}\neq 0$, however for simplicity let us just set  $\mathcal{C}= 0$, this might suggest that the Bekenstein-Hawking entropy comes from the entaglement entropy of the gravitons and matter fields. This can be seen if we rewrite the final equation as
\begin{eqnarray}
  S+2 \pi  \alpha\left(R^2+3\lambda R+3\lambda^2 \right) e^{-\frac{R}{\lambda}}=\frac{A}{4}.
\end{eqnarray}
This equation has the interpretation that the total entropy inside the observable universe is a sum of the entropy of matter and corrections that arise from the graviton-matter entanglement entropy. In general, we get that at the horizon if we take $\lambda \sim R$, we get
\begin{eqnarray}
   S+\frac{14 \pi  \alpha R^2}{e} = \pi R^2.
\end{eqnarray}

\subsection{A connection with the Padmanabhan conjecture}
Further, as we shall argue, it is not difficult to observe from the last equations that one can basically obtain a similar expression to the Padmanabhan conjecture according to which the cosmic space emerges as the cosmic time progress and link the emergence of space to the difference between the number of degrees of freedom on the boundary and in the bulk \cite{33}. We, therefore, associate to $A$ a number of boundary degrees of freedom $N_{\rm boundary} $, while to the second term a number of bulk degrees of freedom $N_{\rm bulk} $, namely we get
\begin{eqnarray}
     \underbrace{\frac{A}{4}}_{N_{\rm boundary} } = \underbrace{2 \pi  \alpha\left(R^2+3\lambda R+3\lambda^2 \right) e^{-\frac{R}{\lambda}}+S}_{N_{\rm bulk}}.
\end{eqnarray}
where it was assumed $\mathcal{C}=0$. This implies that
\begin{eqnarray}
    N_{\rm boundary}=N_{\rm bulk}
\end{eqnarray}
as pointed out by Padmanabhan \cite{33}. However the most general case is to take $\mathcal{C} \neq 0$, in such a case we have 
\begin{equation}
     \underbrace{\frac{A}{4}+\mathcal{C}}_{N_{\rm boundary} } - \underbrace{2 \pi  \alpha\left(R^2+3\lambda R+3\lambda^2 \right) e^{-\frac{R}{\lambda}}+S}_{N_{\rm bulk}} \geq 0
\end{equation}
implying that a nonzero $\mathcal{C}$ indicates deviations from $N_{\rm boundary}=N_{\rm bulk}$. Basically this difference might drive the expansion of the universe which according to Padmanabhan reads
\begin{eqnarray}
       N_{\rm boundary}-N_{\rm bulk}\geq 0. 
\end{eqnarray}
Such modification to the Padmanabhan's idea were suggested in \cite{J4}, while this difference can explain the emergence of space as time progresses
\begin{eqnarray}
       \frac{\Delta V}{\Delta t}=l_p^2 \left(N_{\rm boundary}-N_{\rm bulk}\right)\geq 0. 
\end{eqnarray}
In that sense $\mathcal{C}$ could be related to the entropy of the space-time itself that has just been created. As new space emerges, we can further say new particles (spacetime quanta) have also emerged and as a consequence  we can associate a chemical potential contribution $\mu$ such that $\delta \mathcal{C}=\mu \delta N_{\rm quanta}$, where $N_{\rm quanta}$ gives the number of new particles. The Hawking-Bekenstein entropy then reads
\begin{eqnarray}
    \delta S_{\rm total}=\frac{\delta A}{4}+\mu\, \delta N_{\rm quanta}.
\end{eqnarray}
To conclude, we have argued that modifications to entropy lead to modifications in Newton's law of gravity, effectively resulting in a modified gravity law. The correction term within the entropy expression is obtained from the contribution due to gravitons with matter fields aligning with the Hawking-Bekenstein entropy at the horizon. This suggests that $\alpha$ emerges from the entanglement entropy of gravitons inside the universe. The phenomenological implication of such corrections in large distances, as we shall show in the next section lead to the modification of the Einstein field equations which can explain the phenomena of dark matter in the universe.

\subsection{Recovering Einstein's field equations}
\textbf{First approach.} Using this insight, let's see if we can recover the Einstein field equations. We will assume the validity of the holographic principle and the notion that each unit of information occupies a single bit cell. This allows us to express \cite{Verlinde:2016toy}
\begin{equation}
N = \frac{A}{l_p^2}
\end{equation}
where $l_p^2 = \frac{G\hbar}{c^3}$, we will work in units $G=\hbar=c=1$. In this first approach we consider the modification of the entropy while working with the Newtonian potential. According to statistical mechanics the entropy of a system is proportional to the number of bits. Now using the relation
\begin{equation}
N =4S ~~.
\end{equation}
we get
\begin{eqnarray}
    dN=dA+\alpha \frac{e^{-\frac{\sqrt{A/\pi}}{2\lambda}} \left(2 \sqrt{\pi}\lambda + \sqrt{ A}\right)}{ 2 \sqrt{\pi}\lambda} .dA
    \end{eqnarray}

In other words, this is the bit density on the screen. On the other hand, according to the  equipartition law,  the energy associated with mass $\mathcal{M}$ can be expressed as \cite{Verlinde:2016toy}
\begin{equation}
\label{e20}
\mathcal{M} = \frac{1}{2} \int_{{\cal S}} T ~dN.
\end{equation}
According to Verlinde, the local temperature $T$ on the screen is given by
\begin{equation}
T = \frac{\hbar}{2\pi} n^b \nabla_b \Phi_N,
\end{equation}
where $T$ is measured from infinity. So Eq.~(\ref{e20}) is written as
\begin{equation}
\label{e25}
\mathcal{M}= \frac{1}{4 \pi G} \int_{{\cal S}}~\nabla \Phi_N \left[1+   \alpha  \frac{e^{-\frac{\sqrt{A/\pi}}{2\lambda}} \left(2 \sqrt{\pi}\lambda + \sqrt{ A}\right)}{ 2 \sqrt{\pi}\lambda } \right] \cdot dA.
\end{equation}
The first mass term is the Komar mass, and can be easily shown that
\begin{equation}
\label{e25}
M = \frac{1}{4 \pi G} \int_{{\cal S}}~\nabla \Phi_N \cdot dA.
\end{equation}
From here we get for the total mass
\begin{eqnarray}
\label{e25}
\mathcal{M} &=&M +M_{\alpha}
\end{eqnarray}
where we defined
\begin{equation}
M_{\alpha}=\frac{\alpha}{4 \pi G} \int_{{\cal S}}~\nabla \Phi_N \left[  \frac{e^{-\frac{\sqrt{A/\pi}}{2\lambda}} \left(2 \sqrt{\pi}\lambda + \sqrt{ A}\right)}{ 2 \sqrt{\pi}\lambda } \right] \cdot dA
\end{equation}
which as we shall argue mimics the role of dark matter. The result gives
\begin{equation}
M_{\alpha} = \alpha M \left[ \frac{e^{-\frac{\sqrt{A/\pi}}{2\lambda}} \left(2 \sqrt{\pi}\lambda + \sqrt{ A}\right)}{ 2 \sqrt{\pi}\lambda }\right]
\end{equation}
In the above discussion we have assumed a fixed value for $A$ and $R=R_0$. If we again use $A=4 \pi R_0^2$ and and we approximate $e^{-R_0/\lambda} \to 1/e$ using $R_0 \sim \lambda$, we get
\begin{eqnarray}
     M_{\alpha} \simeq  \frac{\alpha M}{e} + \frac{\alpha M R_0 }{e\,\lambda}
\end{eqnarray}
Using $\beta=\alpha/e$, from the above equations we get
\begin{eqnarray}
   \mathcal{M}= M(1+\beta)+\frac{\beta M R_0 }{ \lambda}.
\end{eqnarray}
On the other hand, the total mass for normal matter can expressed as a volume integral of the stress energy tensor $T_{ab}$ in terms of the relation
\begin{equation}
\label{e27}
M= 2 \int_{\Sigma} \left(T_{ab} - \frac{1}{2} T g_{ab}\right) n^a ~\xi^b ~dV.
\end{equation}
With these results in mind, let us note that we can express the Komar mass in terms of the Ricci tensor $R_{ab}$ and the Killing vector $\xi^a$ \cite{Verlinde:2016toy}. To do so, one needs to apply the Stokes theorem along with the Killing equation for $\xi^a$: $\nabla^a \nabla_a \xi^b = -{R^b}_a\xi^a$. Finally one can get 
\begin{equation}
\mathcal{M}= \frac{1}{4\pi G} \int_{\Sigma} R_{ab}~n^a~\xi^b ~dV ~~.
\end{equation}
It is worth noting that $\Sigma$ represents the three dimensional volume bounded by ${\cal S}$ which is the holographic screen and $n^a$ is the normal. Combining these equations, we can obtain the entropy corrected Einstein's equation as  
\begin{equation}\notag
\int_{\Sigma}\Big[R_{ab} - 8\pi G \Big(T_{ab} - \frac{1}{2}Tg_{ab}\Big)(1+ \beta+\frac{\beta R_0}{\lambda})\Big]n^a~\xi^b~dV=0
\end{equation}
This implies that in a spherically symmetric and static space time we can get the Einstein's equation as 
\begin{equation}
R_{ab}= 8\pi G \Big(T_{ab} - \frac{1}{2} T g_{ab}\Big) \left(1+ \beta+\frac{\beta R_0}{\lambda}\right)
\end{equation}

\textbf{Second approach.} Finally, there is a second approach that can be used to obtain the same result. That is to write the entropy as $S'=A/4$ (see Eq. 80), namely we fix the entropy $N=4S'=4S+8 \pi \alpha\left(R^2+3\lambda R+3\lambda^2 \right) e^{-\frac{R}{\lambda}}=A$ and modify the total mass using the modified potential (Yukawa potential) we get
\begin{equation}
 \mathcal{M} = \frac{1}{4 \pi G} \int_{{\cal S}}~\nabla \Phi \cdot dA=M+\alpha M \frac{R+\lambda}{\lambda}e^{-\frac{R}{\lambda}}.
\end{equation}
We can apply the equation
\begin{equation}
\mathcal{M} = \frac{1}{4\pi G} \int_{\Sigma} R_{ab}~n^a~\xi^b ~dV ~~.
\end{equation}
and for the normal matter can expressed as a volume integral of the stress energy tensor $T_{ab}$ in terms of the relation
\begin{equation}
\label{e27}
M= 2 \int_{\Sigma} \left(T_{ab} - \frac{1}{2} T g_{ab}\right) n^a ~\xi^b ~dV.
\end{equation}
Considering the case $\lambda \sim R_0$ and $e^{-R_0/\lambda} \to 1/e$, and combining these equations, we can obtain the same form for the entropy corrected Einstein's equation as 
\begin{align}
R_{ab}  = 8\pi G \Big(T_{ab} - \frac{1}{2}Tg_{ab}\Big)(1+\alpha/e+\frac{\alpha R_0}{e\,\lambda})
\end{align}
Our pursuit involves two approaches: The first approach hinges on the modification of entropy while engaging with the Newtonian potential. The second approach centers on the modified Newtonian potential while assuming a fixed entropy. 

\subsection{Recovering $\Lambda$CDM cosmological model}
In our analysis we have discovered that the energy-momentum tensor is altered by the additional contribution, effectively mimicking dark matter. This modification of the right-hand side truly arises from the adjustments made to the law of gravity, stemming from the correlation between gravitons and matter fields encapsulated in $\beta$. Comparable modifications to Einstein's equations due to entropic corrections have been derived in previous works (see, for instance, \cite{17}). As an illustrative example, let's consider a flat, homogeneous, and isotropic universe described by the Friedmann-Robertson-Walker (FRW) metric.
\begin{equation}
ds^2=-dt^2+a^2(t)\left[dr^2+r^2(d\theta^2+\sin^2\theta
d\phi^2)\right],
\end{equation}
along with the presence of the cosmological constant (due to dark energy) which is encoded in the energy-momentum part via 
\begin{eqnarray}
    T^{\rm dark\,\, energy}_{ab}=-\frac{\Lambda}{8 \pi G} g_{ab}.
\end{eqnarray}
One can further define $R=a(t)r$ and the dynamical apparent
horizon, a marginally trapped surface with vanishing expansion, is
determined by the relation
\begin{eqnarray}
h^{\mu
\nu}(\partial_{\mu}R)\,(\partial_{\nu}R)=0,
\end{eqnarray}
where $h^{\mu\nu}$ represents the two-dimensional metric. A straightforward calculation yields the apparent horizon radius for the FRW universe as $R=ar=H^{-1}$, where $H=\dot{a}/a$ denotes the Hubble parameter. With this, one can derive the dynamical equation in cosmology. Considering the matter source in the FRW universe, we will assume a perfect fluid described by the stress-energy tensor.
\begin{equation}
T_{\mu\nu}=(\rho+p)u_{\mu}u_{\nu}+pg_{\mu\nu}.
\end{equation}
where $u^{\nu}$ represents the four-velocity of the fluid as observed in the rest frame. Since the fluid is assumed to be at rest in comoving coordinates, the four-velocity is given by $u^{\mu}=\left(1, 0, 0, 0 \right)$. Additionally, considering the continuity equation $\dot{\rho}+3H(\rho+p)=0$, for the case of several matter fluids we get
\begin{eqnarray}
    H^2=\frac{8 \pi G}{3} \sum_i \rho_i \left(1+ \beta_i+\frac{\beta_i}{\lambda H_0}\right),
\end{eqnarray}
where $H_0=1/R_0$. In this case, we expect gravitons to couple with each matter fluid with a different $\beta_i$. If we define $\rho_{\rm crit}=\frac{3}{8 \pi G}H_0^2$, and we use the solutions for densities $\rho_i=\rho_{i 0} a^{-3 (1+\omega_i)}$, along with the parameters $\Omega_i=\Omega_{i0}(1+z)^{3(1+\omega_i)}, \, \Omega_{i0}=  8 \pi G \rho_{i0}/(3H_0^2)$, we get 
\begin{equation}
  E^2(z)=\left(\frac{H}{H_0}\right)^2=\sum_i\Omega_{i}(1+\beta_i)+ \frac{\sum_i\Omega_{i} \beta_i}{\lambda H_0 }.
\end{equation}

We shall elaborate  the physical interpretation of the final term, which simulates the influence of dark matter within the $\Lambda$CDM framework. To achieve a comparable effect to cold dark matter, it is intuitive to explore in more details the expression
\begin{equation}
   \sum_i\Omega_{i} \beta_i= \beta_1\Omega_{\rm B,0}(1+z)^{3}+\beta_2 \Omega_{R,0}(1+z)^4+\beta_3 \Omega_{\Lambda,0},
\end{equation}
where $\beta_{1,2,3}$ denote the coupling parameters. In our framework, regarding the late-time universe, we postulate negligible interaction between gravitons and photons, hence anticipating $\beta_2$ to be exceedingly small or zero. However, the focal point of interest, as we shall elaborate below, lies in the emergence of apparent cold dark matter attributable to modifications in Einstein's field equations. Consequently, dark matter can be construed as an apparent phenomenon, originating from the behavior of baryonic matter. Initially, we observe that for the total matter contribution, we have the baryonic component alongside the apparent dark matter corrections, as given by [incorporating the constant $c$ for coherence],
 \begin{eqnarray}\notag
     \Omega_M(z)&=&\left[\Omega_{B,0}+\beta_1\Omega_{B,0}\left(1+\frac{ c}{\lambda H_0}\right)\right](1+z)^3\\
     &=&\bar{\Omega}_{M,0}(1+z)^3.
 \end{eqnarray}
 For the total radiation component in general we expect a contribution from the coupling between the graviton and radiation as well (which can play important role in the early universe). Thus, we arrive at the following conclusion:
\begin{eqnarray}\notag
      \Omega_R(z)&=&\left[\Omega_{R,0}+\beta_2 \Omega_{R,0}\left(1+\frac{c}{\lambda H_0}\right)\right](1+z)^4\\
      &=&\bar{\Omega}_{R,0}(1+z)^4.
\end{eqnarray}
For the late time universe we expect $\beta_2=0$.  In a similar way, we also expect a contribution from the coupling between the graviton and dark energy 
\begin{eqnarray}\notag
      \Omega_{\Lambda}&=&\left[\Omega_{\Lambda,0}+\beta_3 \Omega_{\Lambda,0}\left(1+ \frac{c}{\lambda H_0}\right)\right](1+z)^0\\
      &=& \bar{\Omega}_{\Lambda,0}(1+z)^0.
\end{eqnarray}
Combining all these equations, we obtain in a compact form
\begin{equation}
  E^2(z)= \bar{\Omega}_{M,0}(1+z)^3+\bar{\Omega}_{R,0}(1+z)^4+\bar{\Omega}_{\Lambda,0},
\end{equation}
An open question to be addressed is to ask: to what extent does the energy-momentum tensor of the gravitational field itself (gravitons) contributes to the dark energy? Given the fact that there exists a relation between $\lambda$ and cosmological constant (see, Eq. (25)), it is natural to expect a contribution to the total amount of dark energy from gravitons, namely using
 \begin{eqnarray}
     \lambda \sim \frac{1}{\sqrt{\Lambda}},
 \end{eqnarray}
but we also know that
 \begin{eqnarray}
     \Omega^{\rm dark\, energy}_{\Lambda,0}=\frac{\Lambda c^2}{3 H_0^2}.
 \end{eqnarray}
Hence one can propose a contribution to dark energy density 
 \begin{eqnarray}
     \Omega^{\rm gravitons}_{\Lambda,0}=\frac{c^2}{\lambda^2 H_0^2}.
 \end{eqnarray}
From these relation we end up with 
 \begin{eqnarray}
  \bar{ \Omega}_{M,0}=\Omega_{B,0}+\beta_1\Omega_{B,0}\left(1+\sqrt{  \Omega^{\rm gravitons}_{\Lambda,0}}\right),
\end{eqnarray}
along with
\begin{eqnarray}
      \bar{\Omega}_{R,0}=\Omega_{R,0}+\beta_2 \Omega_{R,0}\left(1+\sqrt{  \Omega^{\rm gravitons}_{\Lambda,0}}\right), 
\end{eqnarray}
and
\begin{eqnarray}
      \bar{\Omega}_{\Lambda,0}=\Omega_{\Lambda,0}+\beta_3 \Omega_{\Lambda,0}\left(1+ \sqrt{  \Omega^{\rm gravitons}_{\Lambda,0}}\right). 
\end{eqnarray}
We therefore found that not only the baryonic matter but in general the radiation and dark energy are modified due to the interaction with the gravitons. For the late time universe  we expect the coupling $\beta_2 \sim 0$, and if we further take no direct coupling between gravitons and dark energy $\beta_3 \sim 0$, we get only an apparent cold dark matter effect via $\beta_1$. However, there is another possibility to take 
\begin{eqnarray}
    \Omega^{\rm gravitons}_{\Lambda,0} \sim  \Omega_{\Lambda,0}^{\Lambda CDM}=0.7,
\end{eqnarray}
in such a case if we take
\begin{eqnarray}
    \Omega_{B,0}=\Omega_{B,0}^{\Lambda CDM}=0.05,\,\,\,\,\bar{ \Omega}_{M,0}= \Omega_{M,0}^{\Lambda CDM}=0.3,
\end{eqnarray}
yielding
\begin{equation}
\Omega_{M,0}^{\Lambda CDM}=\Omega_{B,0}^{\Lambda CDM}+\Omega_{D,0}^{\Lambda CDM},
\end{equation}
which gives $\beta_1=2.7$.  From the last equation we obtain an alternative expression for the apparent dark matter
\begin{eqnarray}
    \Omega_{D,0}^{\Lambda CDM}=\beta_1 \Omega_{B,0}^{\Lambda CDM}(1+\sqrt{\Omega_{\Lambda,0}^{\Lambda CDM}}).
\end{eqnarray}
This demonstrates that the influence of dark matter primarily arises from the interaction between the graviton and baryonic matter. These findings align closely with those explored in \cite{J1,J2,J3,J4,J5}. In both scenarios, dark matter naturally manifests as an emergent phenomenon. A second crucial observation is that gravitons, to some degree, can assume the role of dark energy. Similar conclusions were drawn, as seen in \cite{Alves:2009sh}, employing Visser's adaptation of general relativity that incorporates graviton mass \cite{Visser:1997hd}. This ultimately contributes to the framework of $\Lambda$CDM.
\begin{equation}
  E^2(z)= \Omega_{M,0}^{\Lambda CDM}(1+z)^3+\Omega_{R,0}^{\Lambda CDM}(1+z)^4+\Omega_{\Lambda,0}^{\Lambda CDM},
\end{equation}
where $\Omega_{R,0}^{\Lambda CDM}=\Omega_{R,0}$ (using $\beta_2=0$) and $\Omega_{\Lambda,0}^{\Lambda CDM}= \Omega^{\rm gravitons}_{\Lambda,0}$ (using $\beta_3=0)$. 

\section{Conclusion}
In the present paper, we investigated the implications of graviton mass  which could arise naturally through a Higgs-like mechanism facilitated by the symmetry breaking of diffeomorphisms. This mechanism holds particular significance as it yields a gravity theory devoid of instabilities.
Another notable advantage lies in the attainment of a graviton mass within the framework of General Relativity, circumventing the fine-tuning issues associated with massive gravity theories. The Higgs-like mechanism that gives mass to gravitons, basically modifies the Newton's law in large distances. Note that this is similar to the the recent idea \cite{J6}, where it was assumed that that the superconductor dark energy medium gives mass to photons, which then modifies the Newton's law. In the present paper however we pointed out that the massive gravitons contribute to the amount of dark energy and, to some extend, they can basically play the role of dark energy. 

It is further argued how the field equation with a cosmological constant can be interpreted as a mass term in the linearized wave equation, namely we argue about the emergence of massive graviton. We then show how the presence of the massive mode
intricately reshapes the grarvitational potential akin to a Yukawa-like potential. Notably, this long-range force modifies Newton's law and this might explain the phenomena of dark matter. In particular, we have shown how the MOND-like acceleration is recovered in this model. Universal relations between dark matter and dark energy in terms of accelerations and force are obtained.   

The most important finding in the present paper is the derivation of the modified law of gravity using the modification of the the Verlindes entropic force relation due to the presence of the massive graviton.  In our case, we pointed out that gravitons indeed wield significant influence, resulting in a modified Newton's law of gravity \cite{18}. By doing so, we may potentially resolve the inconsistencies in Verlinde's entropic force relation, as argued in \cite{Dai:2017qkz}. 

Another important finding lies in the manifestation of Bekenstein-Hawking entropy at the horizon, stemming from the gravitational influence of gravitons. This phenomenon is intricately encoded within the correlation between matter fields and gravitons, characterized by the parameter $\alpha$. This correlation persists when the wavelength of gravitons approaches the scale of the observable radius. Within this observable radius, the entanglement entropy resulting from the interaction between gravitons and matter fields is non-zero, converging to zero precisely at the horizon. This observation further validates Padmanabhan's conjecture, which establishes a connection between the degrees of freedom within the bulk and those at the boundary. Our findings demonstrate the natural emergence of this scenario in our specific context.  

As a culmination of our work, we have derived the corrections to the Einstein field equations, resulting in the formulation of the $\Lambda$CDM model where dark matter emerges as an effect rather than a fundamental component. Our findings prompt a reconsideration of gravity's duality: on one front, it presents itself as an entropic force, rooted in the universal language of entropy; conversely, it exhibits characteristics of a long-range force attributed to the mass of gravitons within the particle framework. Within the realm of entropy and information theory, the graviton is conceptualized as a quasi-particle. The naturally arising modifications to the law of gravity thus provide a coherent framework for explaining dark matter as an apparent effect, alleviating the need for additional assumptions. \color{black}{As a final note we would like to point out that the the full theory in the present paper is not yet complete, for example, one obvious problem is the omission of terms that could have been used to construct a generally coordinate invariant Lagrangian. It is well known that with 4 scalars one can construct a density or measure of integration independent of the metric, see for example \cite{88,89,90,91}. We plan to do further investigations in the near nature.} 
\color{black}

\section*{Acknowledgments}
AFA is grateful to Hassan Alshal for useful discussions. We would like to thank Douglas Singleton for important comments on the manuscript.



\end{document}